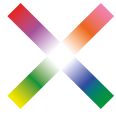

**Dejan Grba**
dejangrba@gmail.com
Artist, researcher, and scholar,
Belgrade, Serbia

# Strange Undercurrents: A Critical Outlook on AI's Cultural Influence

While generative artificial intelligence (generative AI) is being examined extensively, some issues it epitomizes call for more refined scrutiny and deeper contextualization. Besides the lack of nuanced understanding of art's continuously changing character in discussions about generative AI's cultural impact, one of the notably underexplored aspects is the conceptual and ideological substrate of AI science and industry whose attributes generative AI propagates by fostering the integration of diverse modes of AI-powered artmaking into the mainstream culture and economy. Taking the current turmoil around the generative AI as a pretext, this paper summarizes a broader study of AI's influence on art notions focusing on the confluence of certain foundational concepts in computer science and ideological vectors of the AI industry that transfer into art, culture, and society. This influence merges diverse and sometimes inconsistent but somehow coalescing philosophical premises, technical ideas, and political views, many of which have unfavourable overtones.

## 1. Introduction

With the 2022 release of popular online services and tools for text-to-image (TTI) synthesis, such as DALL·E, Leonardo, Midjourney, and Stable Diffusion, and the incorporation of diffusion model routines into offline software, generative artificial intelligence (generative AI) went mainstream. Featuring user-friendly interfaces and streamlined functionality, generative AI systems lowered the technical knowledge barriers for working with multimodal machine learning models that produce high-fidelity output, which expanded the AI's creative user base beyond tech-savvy artists, artistically inclined programmers, and researchers. Amateurs, hobbyists, and enthusiasts as well as professional artists and studios showcase, share, and monetize their generative AI-produced content on social media platforms and portfolio websites. They enter and sometimes win art competitions[1] (Roose 2022; Parshall 2023) and attempt to copyright their visuals (Appel et al. 2023), stirring an increasingly polarizing public debate about generative AI's economic, ethical, and legal consequences.

  By composing prompts as keywords and model directives, a TTI user acts as a task definer and evaluator of the resulting images, and the AI system generates visual concepts and outputs the corresponding pixel arrangements.[2] The limitations of the existing TTI models make it hard to achieve the desired high-quality visual output, so prompting amounts to an iterative trial-and-error process. This expressive challenge to users' diverse notions of visual motifs, styles, mediums,

**1.** Examples include Jason Allen's image *Théâtre D'opéra Spatial* (2022, produced using Midjourney) which won in the digital category at the Colorado State Fair in 2022, and Boris Eldagsen's *Pseudomnesia: The Electrician* (2022, produced using DALLE2) which won the 2023 Sony World Photography Award.

**2.** For an overview of generative AI (multimodal generative models, diffusion models, and TTIs), see Radford et al. (2021), Ho et al. (2022), Ramesh et al. (2022), and Yang et al. (2022).

**3.** See Butler (2003) and Navas et al. (2014).

**4.** The term art brut ("raw art" or "rough art") was introduced in the 1940s by French artist Jean Dubuffet.

**5.** See, for instance, Epstein et al. (2023), McCormack et al. (2023, 3), and Sanchez (2023).

techniques, effects, and other common formal attributes has spurred a burgeoning online scene for sharing prompts, prompting techniques, and prompt-image pairs (on websites such as Prompt Hero) and trading them (on marketplaces such as PromptBase, Promptrr.io, Prompti AI, or PromptScoop).

Since they depend on a predictive amalgamation of styles and other features derived from digital samples of existing media (painting, drawing, photography, and text), TTIs (and generative AI systems more generally) are regarded as sophisticated remediation apparatuses related to, but distinct from style appropriation in postmodernist art and earlier remix cultures (Smith and Cook 2023, 2; Bolter 2023, 195–207).[3] However, despite TTI's disposal of a relatively diverse visual arts corpus, prompting practices largely privilege figurative and descriptive plastic motifs in popular genres of "surreal" or fantasy art, game art, comics, anime, or illustration, with a fixation on surface aesthetics and genre-specific stylistic norms at the expense of other important poetic factors (McCormack et al. 2024). This usage trend parallels the inevitable prevalence of cultural norms in generative AI's training datasets on account of which the TTI imagery often perpetuates and sometimes reinforces stereotypes, biases, and cultural hegemonies (McCormack et al. 2023). For all these reasons, the TTI scene can be considered as a conceptual antipode of art brut—art created by individuals operating beyond the official cultural boundaries (obscure amateurs, psychiatric patients, prisoners, etc.) and distinguished by its uninhibited freshness, non-compliance to expressive canons, and disregard of training-imposed conventions.[4]

While many researchers note that synthetic surface mimicry of popular visual styles does not constitute an artistic innovation and that, at this point, TTIs do not pose a serious threat to human art,[5] they are aggressively pitched as artistic tools. In a broader view, the AI industry's introduction of consumer-grade tools for artmaking and the popularization of other machine learning technologies (e.g. generative adversarial networks) for artistic purposes has never been an innocent or disinterested byproduct of AI's evolution. Releasing attractive devices for creative expression aids the AI industry's marketing, development, and public relations as widely adopted products become "indispensable", provide beta testing feedback and learning data from a large user base, and help associate AI with unique human faculties such as artmaking. The strategy has apparently worked well with the TTIs; uninhibited by the minuscule historical distance, the media, tech-pundits, and some scholars effuse about generative AI's disruptive power over and beyond art. For instance, Lev Manovich (2023) describes generative AI as a revolution comparable in magnitude to the adoption of linear perspective in Western visual arts and the invention of photography. Others believe that generative AI is a profoundly impactful medium whose "synthesis of human intuition and machine capabilities" represents a "paradigm shift" that "heralds a renaissance in artistic expression, offering glimpses into the limitless possibilities that lie ahead in this dynamically evolving art landscape" (Novaković and Guga 2024). They claim that generative AI transcends a mere artistic tool and makes a crucial step toward the fulfilment of the creative industries' long-standing goal to democratize creativity into a more socially integrated and economi-

6. Somewhat paradoxically, such enthusiastic claims tend to ignore the inherent temporal- and context-relativity (instability) of artistic traditions and thus the social porosity of artmaking professions.

7. The paper summarizes a part of the study that examines AI's influences on professional and popular art notions and critiques them within the perspective of the AI's often disturbing techno-cultural underpinnings.

cally productive force and thus redefine the "traditional exclusivity" of artistic roles (Kishor 2023).[6]

Conversely, authors such as Epstein et al. (2023), McCormack et al. (2023; 2024), Sanchez (2023), and contributors to Wilde et al. (2023), have identified and discussed a plethora of generative AI issues and fallouts, offering a more clearheaded approach. Generative AI's most salient problems include the legal and ethical concerns about online data use for model training (data laundering, copyrighted and non-consensual data acquisition, automated appropriation of developed artistic styles), biases (ethnic, racial, gender, cultural), modelling constraints (models as "cultural atoms"), the limitations of text-based paradigm for visual expression, the narrow levels of output and authorial control, the simplistic notions of style (in designing and using models), flimsy aesthetics (derivative visuals, conventionalization, homogenization), expressive novelty and poetic cogency inferior to other artmaking practices, systemic censorship, and short- and long-term impact on the creative and media industries (loss of human skills, job precarity, improved deepfaking, fake news).

The dynamic of opposing sentiments about TTIs' expressive capacities and limitations, as well as their sociopolitical and cultural issues, has turned generative AI into the word frequency star of critical AI studies. However, despite the scope and depth of their findings, critical AI studies largely resonate with academia, while the public and some segments of professional communities are being saturated with hyped-up rhetoric and generalized views that shape the prevailing art notions and directly or indirectly influence artistic practices in a range of fields. In such context, certain aspects of generative AI critique require more refined scrutiny and deeper contextualization. Notably, the discussion about AI's impact on art notions lacks a nuanced understanding of art's continuously changing identity brought about by the modernist avantgardes, postmodernism, and experimental art practices, which both reflects and retrenches the prevailing art dilettantism across the AI science/tech sector and affects the ways of pondering art's natures, functions, and futures. A related topic calling for keener attention is the conceptual and ideological substrate of the computer science and AI industry whose attributes generative AI helps disseminate by facilitating the proliferation of digital artefacts and fostering the integration of computational art into the mainstream culture and economy. In this paper, I focus on that haunting substrate.[7] It merges diverse and sometimes incongruous but somehow coalescing technical concepts, philosophical premises, and political views, many of which have the overtones of alienation, sociopathy, and misanthropy. They are largely obscured in the debates about AI's transformations of art and society and remain underexposed in AI studies so, in the closing section, I outline some of their manifestations in generative AI and introduce several viewpoints for a further critique of AI's cultural zeitgeist.

## 2. Undercurrents

A collection of tendencies and syndromes in the conceptual and ideological undertows of AI science, technology, and industry wields a strong if seemingly indirect influence on cultural mindsets and art notions. It includes the fetishism of machinic agency, the mutual equalization of computers and humans, statistical reductionism, sociotechnical blind-

**8.** For instance, the narcissistic idea that sophisticated nonhuman entities (intelligent robots or angels) would strive to become human and gladly accept all the oddities and costs that come in the package underpins Isaac Asimov's novelette *The Bicentennial Man* (1976) and film *Wings of Desire* (1987, directed by Wim Wenders).

ness, and cyberlibertarianism. Their disparity and, in some cases, apparent awkwardness notwithstanding, these factors amalgamate into a powerful flux.

### 2.1 The Fetishism of Machinic Agency

Although it ranks among the most widely and thoroughly discussed AI issues, anthropomorphism remains pervasive and highly detrimental to both AI science/tech and AI/art intersections. It is an innate psychological tendency to assign human cognitive traits, emotions, intentions, or behavioural features to non-human entities or phenomena (Hutson 2012). Exposing a trans-cultural anthropocentric tenet that humanity is the sine qua non of the universe (Tromble 2020, 5),[8] anthropomorphism has steadily pervaded the foundational concepts, terminology, and notions of intelligence in AI science and industry as well as in popular discourse (Salles et al. 2020). Its main aspects are encapsulated in David Watson's remarks (2019, 432, 434–435):

> *A number of [machine] learning algorithms either deliberately or coincidentally mirror certain aspects of human cognition to varying degrees. In a sense, this is only to be expected. For better or worse, we are our own best source of inspiration when it comes to modelling intelligence. There is nothing especially remarkable or problematic about this. However, issues arise when we begin to take these metaphors and analogies too literally. [...] Algorithms are not "just like us" and the temptation to pretend they are can have profound ethical consequences when they are deployed in high-risk and other sensitive domains. By anthropomorphizing a statistical model, we implicitly grant it a degree of agency that not only overstates its true abilities but robs us of our own autonomy. [...] Algorithms can only exercise their (artificial) agency as a result of a socially constructed context in which we have deliberately outsourced some task to the machine. [...] The central point—one as obvious as it is frequently overlooked—is that it is always humans who choose whether or not to abdicate this authority, to empower some piece of technology to intervene on our behalf. It would be a mistake to presume that this transfer of authority involves a simultaneous absolution of responsibility. [...] The temptation to grant algorithms decision-making authority in socially sensitive applications threatens to undermine our ability to hold powerful individuals and groups accountable for their technologically mediated actions.*

Anthropomorphism can be difficult to identify, especially in metaphors where it most frequently appears, which often has undesired consequences (Curry 2023, 178). As Kieran Browne and Ben Swift pointed out (2019, 3), in the language of AI, assertions that a machine "learned", "discovered", "outsmarted", etc. presuppose agency and often imply consciousness but even placing a machine as the subject of a sentence is dubious and deserves examination. The continuous illusionism of intelligent communication or "banal deception" throughout AI's history (Natale 2021) opens a perspective for understanding anthropomorphism and autonomous AI fetishism not just as the side-effects of our evolved bias toward detecting agency, but also in the light of human propensity for deception and self-deception (Trivers 2011).

Hence, it is often hard to evaluate, and easy to dismiss, the difference between the effectiveness of human intelligence and the efficiency

**9.** Apocryphally related to a Victorian parlor game (Athanasius 2019), The Imitation Game involves a human evaluator who judges a natural language conversation between another human and a machine designed to generate human-like responses. Participants converse through a text-only channel (written messages) and the evaluator knows that one of the two conversationalists is a machine. The machine passes the test only if the evaluator cannot reliably distinguish between the machine's and human's messages after a fixed period.

of specialized artificial processes related to our concepts of intelligence. Of course, there is no reason nor justification for conflating a non-living system with a biological entity just because both can exhibit some behaviours and perform certain functions that are computationally interpretable. Nevertheless, the media and some AI scientists repeatedly associate the performance of state-of-the-art machine learning systems with human cognitive traits such as intuitive physics, intuitive biology, intuitive psychology, causal models, active social learning, conceptualization, subconscious abstraction, generalization, analogy-making, and common-sense reasoning—the very capabilities they lack the most (Mitchel 2019, 140, 195–199). Throughout the history of computer science, the epistemological and metaphysical confusions caused by conflating human intelligence and machine performance have rendered anthropomorphism and AI inseparable, and some authors suggest that it is more feasible to manage anthropomorphism in AI research than purge it (Proudfoot 2011; Watson 2019, 417–440). In this light, we can view AI as an important part of techno-cultural and social dynamics in which a what becomes a who and vice versa (Bratton and Agüera y Arcas 2022). Its many problems arise from the awkward understanding of computers vis-à-vis human beings and paradoxical tendencies toward their mutual equalization reaching back to the foundations of computer science and AI.

### 2.2 Computers = Humans

One of the unfortunate consequences of Alan Turing's legacy is the intentional or accidental provision of a "scientific basis" for radical anthropomorphism—viewing and treating human beings as computers. In his paper *On Computable Numbers, With an Application to the Entscheidungsproblem*, Turing first described an "automatic machine", which was later named Turing machine and became one of the key concepts in computer science. The paper was published in 1936, before the advent of automatic computing, when many people in business, government, and research establishments professionally carried out numerical calculations. These human calculators were called "computers" and Turing reemphasized in various forms that the terms "computation" and "computable" in his paper refer to an idealized description of their work (Copeland 2020). Thus, Turing's analogizing of a set of highly structured operations performed by human beings with idealized computing machines makes sense only within the specific historical and utilitarian contexts of his writing. But he ostensibly went from connecting the isolated features of human and machine computation toward conflating human beings with computing machines. In a 1950 paper *Computing Machinery and Intelligence*, Turing proposed the Imitation Game as a method for testing a computational machine's ability to exhibit intelligent behaviour equivalent to, or indistinguishable from, a human. The proposal became known as the Turing Test, and this title has often been used to indicate other behavioural tests for the presence of mind or intelligence in artificial systems.[9] However, Turing cantered his proposal around an unclearly defined concept of intelligence and left many other parts of the discussion open to interpretation, which resulted in a long-lasting controversy (Oppy and Dowe 2020).

Strong objections to the Turing Test posit that with the Imitation Game Turing aimed to legitimize the "null hypothesis" of no behaviour-al difference between certain machines and humans, and that such a perspective is arrogant and parochial because it assumes that we can understand human cognition without first obtaining a firm grasp of its basic principles (Searle 1980; Block 1981). Furthermore, assessing human intelligence through a single, highly formalized layer of linguistic communication (written text) is too narrow to be decisive as thinking is frequently nonverbal and combines verbal and nonverbal mental processes with numerous other factors (Tulio 2021). Turing's flirting with the "null hypothesis" also provides grounds for an argument that aloofness, narcissism, and psychological issues evident throughout his life "conspired" to elicit a misanthropic bitterness, which motivated the infantile computer-human analogy.[10]

The range, character, and persistence of grotesque notions in AI research indicate both conceptual and mental issues, so it is important to acknowledge their connotations and consequences:

> *The separation between "reasonable" and "unreasonable" ideas [in AI science], which we might call superstition is less clear than one might expect. In Computing Machinery and Intelligence, Alan Turing considers the use of a "telepathy-proof room" to protect the integrity of his Imitation Game from players exhibiting extrasensory perception. This may cause us to cringe in hindsight—it's uncomfortable to imagine heroes of science believing such unlikely things. But good science demands open-mindedness and the courage to challenge accepted truths. AI researchers are in a difficult position, expected to dismiss "silly" ideas like telepathy and yet take seriously the idea that bits of metal and silicon might become intelligent if you program them the right way. (Browne and Swift 2019, 2)*

Despite the intellectual challenges of cutting-edge thinking, the leniency in the AI community toward its founders' anthropomorphic tendencies[11] and its members' other quirks is regressive and irresponsible. The notion of personified computers awards the machine (a non-living entity) the role of the Other, places it into our circle of empathy (Singer 2011), and assigns it elevated rights while we have been long surrounded with living "candidates" for expanding our empathy or improving our ethics but still don't treat them consistently and justly: other human beings, animals, and plants. The consequently perverse logic of personified AI implies that we need to devise value systems (urgently needed but inadequately applied to many existing beings) on a purely speculative model of sentient AI.

How do such ambiguities translate to the reality in which the rapid industrialization and widespread application of AI technologies bring about the concentration of wealth and political power that leads to a society contingent on corporate AI interests?

### 2.3 The Autonomous AI Myth

Institutions and relations that involve frequent information exchange and processing can, for some practical purposes, be envisioned and treated as data structures. Thus, quantization, data collection, behavioural tracking, predictive modelling, and various types of decision-making manipulation have long been essential strategies for large-scale information-dependent systems such as governments, industry, marketing, finance, insurance, media, and advertising. The corporate

---

**10.** While acknowledging the caveats of retrospective diagnoses and the subjective nature of diagnosing in general, Henry O'Connell and Michael Fitzgerald's (2003) analysis of Turing's biography and contemporaneous accounts concludes that he met Gillberg, ICD-10, and DSM-IV criteria for Asperger's syndrome, which places him within the autism spectrum disorder.

**11.** Another pioneering giant of computer science and AI, Marvin Minsky, was a proponent of the concept of a computer as a person (see Elis 2014). Jaron Lanier's autobiography *Dawn of the New Everything* (2017) provides several vivid accounts that illustrate the prevalence of such a mindset among Silicon Valley hackers.

**12.** Since its launch in 2005, Amazon's Mechanical Turk has been the largest and most widely known microlabour platform (Mitchell 2019, 84–85). Other platforms include Fiverr, Microworkers, Clickworker, Upwork, TaskRabbit, WorkMarket, Catalant Technologies, Inc., and Toloka.

AI sector increases the extent and intricacy of these strategies by combining massive digital datafication with sophisticated statistical algorithms for profiteering or social engineering, which has many undesirable effects (O'Neil 2016; Zuboff 2019). Statistical reductionism is not exclusive to businesses and can be radicalized by state regimes that deploy AI for authoritarian societal control and governance. For instance, the Social Credit System and the "innovative development pilot zones", implemented by the Chinese government and AI industry in 2014 and 2019 respectively, are based on a state-wide networked surveillance and assessment of citizens' social and business activities with practical repercussions such as the availability of jobs, education, bank loans, electronic services, transportation, and travel (Yang 2022).

While promising improved discovery outcomes in science, increased economic productivity, commodity spectrums, and profits, the AI industry generates problems that affect various demographic groups. They mostly arise from the disparities between its business priorities (maximizing profit/wealth and competitive power), the social impact of its products, and broader societal interests. Since Google's data harvesting operations that started in the mid-2000s and ImageNet's popularization of image scraping practices in the mid-2010s, modern AI development rides a razor-thin line between research and commerce, and the AI industry often abuses it. Programmatically collecting vast amounts of data and using questionable labour practices to assemble it into the model-training datasets (public or private) is ethically dubious, even in the academic context. However, once the economy springs up around such practices, they become harder to control and regulate.

In aggregate, these trends contribute to an illusion that human-created and human-dependent AI systems have high levels of material abstraction and functional autonomy. Pervading both professional and public discourse, the myth of autonomous AI continues the tradition of using human beings as hidden micro-components in large computational architectures since the late 19th century. It has been identified as AI's "sociotechnical blindness" (Johnson and Verdicchio 2017), "fauxtomation" (Taylor 2018), "ghost work" (Gray and Suri 2019), and "human in the loop" complex (Paulsen 2020). The "synergy" of human work extraction and transparency is notoriously evident on largely unregulated online marketplaces for crowdsourced labour (also called microlabour or crowdlabour).[12] A plethora of unethical HR management practices and widespread workforce exploitation on microlabour platforms through a combination of technical features and legal loopholes has been thoroughly documented (Irani and Silberman 2013; Lorusso 2020; Zukalova 2020). With generative AI, algorithmized human labour demands have somewhat changed from their role in trailblazing AI development techniques and pipelines (Williams et al. 2022), but remain enormous and exploitative: many essential tasks are repetitive and meaningless, and labour conditions are precarious and surveilled (Dzieza 2023; Solaiman et al. 2023; GlobalData 2023).

**13.** It is not clear whether or how significantly modern AI's social politics has diverged from the foundational principles of cybernetics in the 1950s and 1960s. In an anticipation of computers thoroughly integrated into human affairs, cybernetics founder Norbert Wiener (1988) criticized control-hungry sciences and technologies of the past and (arguably) strived to empower society by the humane application of bio-inspired, self-regulating artificial systems. However, authors such as Donna Haraway (2016) and Andreas Broeckmann (2016, 113–115, and passim) claim that the technological and biopolitical paradigm of cybernetics related humans to machines ambiguously and was ideologically geared at subjecting humans.

**14.** For a substantial critique of Rand, see the RationalWiki article (2023).

## 2.4 Cyberlibertarianism

Constructing elaborate illusions of autonomous automation and designing labour maximization algorithms are certainly not AI's most impressive achievements but are emblematic of corporate AI's social politics (Crawford 2021, 48–49, 53–87).[13] Since the mid-1960s, the worldviews in computer science communities and IT industries, particularly in the US and other anglophone countries, have been shaped by a bizarre ideological conglomerate of contradictory doctrines, such as utopianism, counterculture, individualism, libertarianism, and neoliberal economics (Turner 2008; Gere 2008; Rushkoff 2022). This ideological assemblage, also called the Californian ideology (Barbrook and Cameron 2008) and cyberlibertarianism (Winner 1997), comprises ideas fuelled by the zeal for technologically mediated lifestyles and future visions with libertarian notions of freedom, social life, economics, and politics. It promotes technological determinism, radical individualism, a deregulated market economy, trust in the power of business, and disdain for the role of government. These values fully make sense only within the context of the right-wing political milieu (Payne 2013; Armistead 2016) and, openly or tacitly, many cyberlibertarians endorse the unblushing egoism promoted by Ayn Rand's dilettante philosophy (Objectivism) but conveniently overlook its bleak sociopathy (McGinnis 2012; Robephiles 2022), sometimes with the overt cynicism of providing the "philosophical authority" for socioeconomic views steeped in technocracy, greed, and exploitation.[14]

Cyberlibertarian tendency to conflate social and political with technical problems can be summarized in the three assumptions of technological manifest destiny: 1. technology is apolitical so it will automatically lead to good outcomes for everyone; 2. new technologies should be deployed as quickly as possible, even with incomplete knowledge about their functioning and societal impacts; 3. the past is generally uninteresting and history has nothing to teach us (Mickens 2018). After the introduction of blockchain technologies in the late 2000s, the cyberlibertarian techno-solutionist politics has been radicalized by the burgeoning start-up mentalities of predominantly white male crypto entrepreneurs obsessed with quick success and tending toward sexism, racism, misogyny, homophobia, and transphobia (UNESCO 2020). Cyberlibertarianism thrives behind the AI industry's facade of objectivity, rationality, progress, and political correctness whereas its reality is dominated by aggressive competitiveness within an adversarial business culture that promotes the most unpardonable tenet of capitalism: prioritizing profit over people (Wiener 2020). AI industry values "uniquely human" skills such as attention, care, critical judgment, taste, imagination, improvisation, spontaneity, sincerity, empathy, intimacy, and humour not because they evidence individuality or authenticity but primarily because they cannot be automated for generating surplus value (Horning 2015; Gosse 2020). In that context, generative AI can be seen as the forefront of the reiterative entrepreneurial process toward emancipating capital from humanity (Dyer-Witheford et al. 2019, 7), in which human work and data provision build systems that automate certain tasks and reconfigure human working and data provision roles in the next iteration.

While some authors deem such logic morally untenable and AI's labour displacement effects destructive in the long term (Eubanks 2018) and others remain undecided (Epstein et al. 2023, 8–11) or claim the **15.** Golumbia also monetizes his essays behind the paywall on Medium (founded by Evan Williams, a tech billionaire, co-founder, and former CEO of Twitter), while Marx hosts his podcast Tech Won't Save Us on YouTube and monetizes it through Patreon.

opposite (Kalish and Wolf 2023), it is worth remembering that, insofar as we take advantage of AI's sociotechnical regime, we share a degree of responsibility for its existence and consequences. This entanglement is evident in the ethical inconsistencies of some leading critics of cyberlibertarianism who selectively enjoy certain layers of its gravy train by patronizing convenient businesses that epitomize the most acute points of their critique, which may be interpreted as unprincipled or hypocritical. For instance, authors such as Shoshana Zuboff, David Golumbia, and Paris Marx choose publishers who sell their books on Amazon.com[15] rather than less lucrative alternatives, such as the Institute of Network Cultures (INC), which allows readers to either purchase INC books on their website or download them for free.

### 3. Conclusion

Although the historical, philosophical, and sociological studies of computer science and AI have explored most of these issues, they require wider attention in artistic communities because the AI industry's instrumentalization of art and creative expression for the promotion of its products serves as one of the high-bandwidth channels for the cultural normalization of questionable presumptions, concepts, economic interests, and political views in its background. The confluence of AI's problematic undercurrents hijacks our cultural intuition (Pedwell 2022), translates into art practices and their public reception (Lossin 2022), and influences the notions of art and creativity in the professional and popular art discourse.

For instance, the claims of generative AI's Promethean role in "democratizing artmaking" reverberate the cyberlibertarian myths about the democratizing powers of markets and digital technologies (Golumbia 2016). They also support the info-capitalist exploitation of creativity (Reckwitz 2017). Anthropomorphism in AI art, its media representation, and public interpretation articulate motives for relegating creative decision-making to AI systems, hedging or minimizing artistic responsibilities, and foregrounding the benefits of automated cultural production (Browne 2022). The resulting notions of art made by autonomous AI entities reinforce the AI industry's sociotechnical blindness. Users' compliance with generative models' censorship criteria (Riccio et al. 2022) upholds the AI industry's confinement of clients' socioeconomic benefits from leveraging its products. Similarly, artists' apparently sensible adoption of first-aid tech solutions against the misappropriation of their work for generative models training, such as data poisoning or style masking (Shan et al. 2023), inadvertently plays in tune with the techno-solutionist rhetoric whereby only the tech (but not the regulation of techno-economic power) can save us (Morozov 2013) and diminishes the vitality of art as a human faculty. AI's troubles get additionally legitimized through the tech science and industry's implicit sanctioning of their creative employees' relational deficiencies and psychological disorders as acceptable trade-offs of otherwise desirable talents (Dayan 2017; Wayne Meade et al. 2018) and through the corporate "justification" of sociopathic entrepreneurs due to the successes of their daring but morally dubious business ventures (Jacoby 2020; Marx 2023).

The sinister undertows of AI-influenced culture can be critiqued further as an amalgamation of economic interests (Golumbia 2009), self-indulgent anthropocentrism (Zeilinger 2021), psychological mech-anisms of self-deception and cognitive compartmentalization (Trivers 2011), as well as virtue-signalling, competitiveness, and exploitative drives (Miller 2019). The fact that shady motives and unflattering features of human nature remain insufficiently considered in AI studies may help explain the ease with which (mis)anthropic contradictions infuse the artworld's and popular notions about artmaking, sometimes with detrimental effects (Grba 2022; 2023). In this paper, I sketched the main aspects of AI's disturbing undercurrents aiming to expand the repertoire of viewpoints for appraising art and creativity in the age of AI. By looking critically into the mise-en-scène of AI's cultural sway, we can cultivate an informed and responsible approach that adds a touch of scepticism when asking how profoundly technological trends, such as generative AI, transform our relationships with art and in which directions they stir the arts' social, economic, and political roles.